\documentclass[prb,aps,twocolumn,floatfix]{revtex4}
\usepackage{graphicx,bm}
\usepackage{subfigure}
\usepackage{comment}
\usepackage{amssymb}
\usepackage{verbatim}

\newcommand{\avg}[1]{\langle #1 \rangle}

\newcommand{\nn}{\nonumber\\}
\newcommand{\up}{\uparrow}
\newcommand{\down}{\downarrow}

\newcommand{\Ra}{\Rightarrow}

\newcommand{\midb}[1]{\left[ #1 \right]}
\newcommand{\smlb}[1]{\left( #1 \right)}
\newcommand{\abs}[1]{\left| #1 \right|}

\newcommand{\quarter}{{1\over 4}}
\begin{document}
\pagestyle{plain}
\title{Spin Transfer Torque for Continuously Variable Magnetization}
\author{Jiang Xiao and A. Zangwill}
\affiliation{School of Physics, Georgia Institute of
Technology, Atlanta, GA 30332-0430}
\author{M. D. Stiles}
\affiliation{Electron Physics Group, National Institute of Standards and
Technology, Gaithersburg, MD 20899-8412}
\begin{abstract}
    We report quantum and semi-classical calculations of spin current
    and spin-transfer torque in a free-electron Stoner model for
    systems where the magnetization varies continuously in one
    dimension. Analytic results are obtained for an infinite spin
    spiral and numerical results are obtained for realistic domain
    wall profiles.  The adiabatic limit describes conduction electron
    spins that follow the sum of the exchange field and an effective,
    velocity-dependent
    field produced by the gradient of the magnetization in the
    wall. Non-adiabatic effects arise for short domain walls but their
    magnitude decreases exponentially as the wall width increases.
    Our results cast doubt on the existence of a recently proposed
    non-adiabatic contribution to the spin-transfer torque due to spin
    flip scattering.
\end{abstract}
\date{\today} \maketitle
\section{Introduction}
The use of electric current to control magnetization in
nanometer-sized structures is a major theme in the maturing field of
spintronics. \cite{Fert:2003} An outstanding example is the
theoretical prediction of current-induced magnetization precession and
switching in single domain multilayers \cite{SB} and its subsequent
experimental confirmation in spin-valve
nanopillars. \cite{Kiselev:2003} The physics of this spin-transfer
effect is that a single domain ferromagnet feels a torque because it
absorbs the component of an incident spin current that is polarized
transverse to its magnetization. The same idea generalizes to systems
with continuously non-uniform magnetization. \cite{Berger:1978,
Bazaliy:1998} This realization has generated a flurry of experimental
\cite{expt} and theoretical work \cite{Waintal:2004, Tatara:2004,
Zhang:2004, Thiaville:2005, Barnes:2005} focused on current-driven
motion of domain walls in magnetic thin films.

The experiments cited just above employ N\'eel-type domain walls with
widths $w\approx 100$ nm. This length is very large compared to the
characteristic length scales of the processes that determine the local
torque.\cite{Stiles:2002,Zhang:2002} Therefore, it is appropriate to
adopt an adiabatic approximation where the spin current is assumed to
lie parallel to the local magnetization. \cite{Berger:1978,
Bazaliy:1998} Surprisingly, the adiabatic prediction for the current
dependence of the domain wall velocity \cite{Berger:1986, Tatara:2004,
Thiaville:2004} agrees very poorly with experiment.  This has led
theorists \cite{Waintal:2004, Tatara:2004, Zhang:2004, Thiaville:2005,
Barnes:2005} to consider non-adiabatic effects and experimenters
\cite{Klaui:2005, Ravelosona:2005} to study systems with domain wall
widths that are much shorter ($w\approx 10$ nm) than those studied
previously.

Two groups \cite{Zhang:2004, Thiaville:2004} have studied the effect
on domain wall motion of a distributed spin-transfer torque
represented by a sum of gradients of the local magnetization with
constant coefficients. For a one-dimensional magnetization ${\bf
M}(x)$, the torque function can be written in terms of two vectors
perpendicular to the magnetization
\begin{equation}\label{lt}
    {\bf N}_{\rm st}(x) = c_1\partial_x\hat{\bf M}  
+ c_2\hat{\bf M}\times\partial_x\hat{\bf M}.
\end{equation}
In general, the coefficients $c_1$ and $c_2$ are functions of position.
The well-established adiabatic piece of the torque is the first term
in Eq.~(\ref{lt}) with a constant coefficient.  Consistent with usage
in the literature, we call all deviations from the adiabatic torque
non-adiabatic.  Any contributions of the second term are then called
non-adiabatic. Zhang and Li \cite{Zhang:2004} derive a contribution
along this second direction in
Eq.~(\ref{lt}) from a consideration of magnetization relaxation due to
spin-flip scattering in the context of an s-d exchange model of a
ferromagnet. \cite{Berger:1986,Zhang:2002} Their arguments lead them
to the estimate $c_2/c_1\approx 0.01$. The authors of
Ref.~\onlinecite{Thiaville:2005} report that a similar value of
$c_2/c_1$ produces agreement with experiment when Eq.~(\ref{lt}) is
used in micromagnetic simulations.

In this paper, we study the applicability of Eq.~(\ref{lt}) to a
free-electron Stoner model with one-dimensional magnetization
distributions of the form
\begin{equation}
    \label{eqn:magnetization}
    {\bf M}(x) = M_s\smlb{\sin\theta\cos\phi, \sin\theta\sin\phi, \cos\theta}.
\end{equation}
In Eq.~(\ref{eqn:magnetization}), the polar angle $\theta(x)$ is
measured from the positive $z$ axis, $\phi(x)$ is the azimuthal angle
in the $x$-$y$ plane, and $M_s$ is the saturation magnetization.  We begin
with the spin current and spin torque for an infinite spin spiral with
constant pitch. This system turns out to be perfectly adiabatic; the
torque is described by Eq.~(\ref{lt}) with $c_2=0$. The same is true
for realistic domain walls of the sort usually encountered in
experiment. Non-adiabatic effects appear only for very narrow
walls. In that case, the torque is non-local and cannot be written in
the form Eq.~(\ref{lt}). The non-adiabatic torque decreases
exponentially as the wall width increases for all realistic domain
wall profiles.  Finally,  our analysis casts doubt on the
existence of a non-adiabatic contribution to the torque due to
spin-flip scattering proposed recently by Zhang and
Li.\cite{Zhang:2004}.  

The remainder of the paper is organized as
follows. Section~\ref{models} describes our Stoner model and the
methods we use to calculate the spin current and spin-transfer
torque. Section~\ref{spinspiral} reports our results for an infinite
spin spiral and Section~\ref{domainwalls} does the same for
one-dimensional domain walls that connect two regions of uniform
magnetization.  Section~\ref{other} relates these calculations to
previous work by others.  Section~\ref{scattering} discusses the
effects of scattering. We summarize our results and offer some
conclusions in Section~\ref{sumcon}.  Two appendices provides some
technical details omitted in the main body of the paper.

\section{Model \& Methods}
\label{models}
The free electron Stoner model provides a first approximation to the
electronic structure of an itinerant ferromagnet. The Hamiltonian is
\begin{equation}
\label{eqn:Hamiltonian}
H=-{\hbar^2\over 2m}\nabla^2 - \mu \bm{\sigma} \cdot {\bf B}_{\rm ex}(x),
\end{equation}
where $\bm{\sigma}=(\sigma_x,\sigma_y,\sigma_z)$ is a vector composed
of the three Pauli matrices and $\mu=g\hbar e/2mc$. The magnetic field
${\bf B}_{\rm ex}(x)$ is everywhere parallel to ${\bf M}(x)$ but has a
constant magnitude.\cite{MagSign} That magnitude is chosen so the
Zeeman splitting between the majority and minority spins bands
reproduces the quantum mechanical exchange energy in the limit of
uniform magnetization:
\begin{equation}
\label{eqn:exchange}
E_{\rm ex}=2\mu |{\bf B}_{\rm ex}|=\hbar^2 k_B^2/m.
\end{equation}
If $E_F=\hbar^2 k_F^2/2m$ is the Fermi energy, the constant $k_B$ in
Eq.~(\ref{eqn:exchange}) fixes the Fermi wave vectors for up and down
spins, $k_F^+$ and $k_F^-$, from
\begin{equation}
\label{eqn:kpm}
k_F^\pm = \sqrt{k_F^2\pm k_B^2}.
\end{equation}

Given ${\bf B}_{\rm ex}(x)$, we use both quantum mechanics and a
semi-classical approximation to calculate the spin accumulation, spin
current density, and spin-transfer torque. The building blocks are the
single-particle spin density ${\bf s}_\pm(x,k_x)$ and the
single-particle spin current density\cite{Qnote} ${\bf Q}_\pm(x,k_x)$
for an up/down ($\pm$) electron with longitudinal wave vector $k_x$.

Summing over all electrons and using the relaxation time approximation
gives the non-equilibrium majority and minority spin density ${\bf
s}_\pm(x)$ and spin current density ${\bf Q}_\pm(x)$ in the presence
of an electric field $E\hat{\bf x}$:
\begin{eqnarray}
\label{eqn:spinandcurrent}
{\bf s}_\pm(x) 
    &=& \int\midb{f_\pm({\bf k}-{eE\tau\over\hbar}\bm{\hat{x}})-f_\pm({\bf k})}{\bf s}_\pm(x,k_x)d^3{\bf k} \nonumber \\
& & \\
{\bf Q}_\pm(x) 
    &=& \int\midb{f_\pm({\bf k}-{eE\tau\over\hbar}\bm{\hat{x}})-f_\pm({\bf k})}{\bf Q}_\pm(x,k_x)d^3{\bf k} \nonumber
\end{eqnarray}
Our use of the function $f_\pm({\bf k})=\Theta(k_F^\pm-\abs{\bf k})$
implies that the distribution of electrons outside the region of
inhomogeneous magnetization are characteristic of the zero-temperature
bulk.\cite{linearBoltzmann} We shall expand this point and comment on
the general correctness of (\ref{eqn:spinandcurrent}) in
Sec.~\ref{scattering}.

The sum ${\bf s}(x)={\bf s}_+(x)+{\bf s}_-(x)$ is the total spin
accumulation (spin density) and ${\bf Q}(x)={\bf Q}_+(x)+{\bf Q}_-(x)$
is the total spin current density. Finally, the distributed spin
transfer torque is\cite{Stiles:2002}
\begin{equation}
\label{eqn:spt}
{\bf N}_{\rm st}(x)=-\partial_x{\bf Q}(x).
\end{equation}
The adiabatic approximation\cite{Berger:1978} to the spin dynamics
leads to a spin current density that is proportional to the local
magnetization, ${\bf Q}_{\rm ad}(x) \propto {\bf M}(x)$. This means
that
\begin{equation}
\label{eqn:adia}
{\bf N}_{\rm ad}(x)  \propto \partial_x {\bf M}(x).
\end{equation}
A main goal of this paper is to study the extent to which the
spin-transfer torque associated with real domain wall configurations
satisfies Eq.~(\ref{eqn:adia}).

\subsection{Quantum}
\label{sec:quantum}

In light of Eq.~(\ref{eqn:magnetization}), the exchange magnetic field
that the enters the Hamiltonian in Eq.~(\ref{eqn:Hamiltonian}) is
\begin{equation}
\label{eqn:exchangeB}
{\bf B}_{\rm  ex}(x) =B_{\rm ex}\smlb{\sin\theta\cos\phi, \sin\theta\sin\phi, \cos\theta}.
\end{equation}
For a given energy, the eigenfunctions for a conduction electron with
wave-vector ${\bf k}$ take the form $\Psi_\pm({\bf r},{\bf
k})=\psi_\pm(x, k_x)e^{ik_y y+ ik_z z}$ where the spinor
$\psi_\pm(x,k_x)$ satisfies
\begin{equation}
    \midb{-{d^2\over dx^2} - k_B^2\smlb{
    \begin{array}[c]{cc}
    \cos\theta & e^{-i\phi}\sin\theta \\
    e^{i\phi}\sin\theta & -\cos\theta 
    \end{array} } }\psi_\pm
    = \kappa_\pm^2\psi_\pm.
    \label{eqn:Schrodinger}
\end{equation}
In this expression,
\begin{equation}
\label{eqn:kdef}
\kappa_\pm^2=k_x^2\mp k_B^2,
\end{equation}
and $\pm$ refers to majority/minority band electrons. The
single-electron spin density and spin current density
are\cite{Stiles:2002}
\begin{equation}
\label{eqn:sesd}
{\bf s}_\pm(x,k_x)={\hbar \over 2}\sum_{\alpha,\beta}
\psi^\ast_{\pm,\alpha}(x,k_x)\bm{\sigma}_{\alpha,\beta}\psi_{\pm,\beta}(x,k_x) 
\end{equation}
and
\begin{equation}
\label{eqn:sescd}
{\bf Q}_\pm(x,k_x)=-{\hbar^2 \over 2m}\sum_{\alpha,\beta} {\rm Im} [
  \psi^\ast_{\pm,\alpha}(x,k_x)\bm{\sigma}_{\alpha,\beta}{d \over dx}
  \psi_{\pm,\beta}(x,k_x)]. 
\end{equation}

\begin{figure}
  \includegraphics[width=8cm]{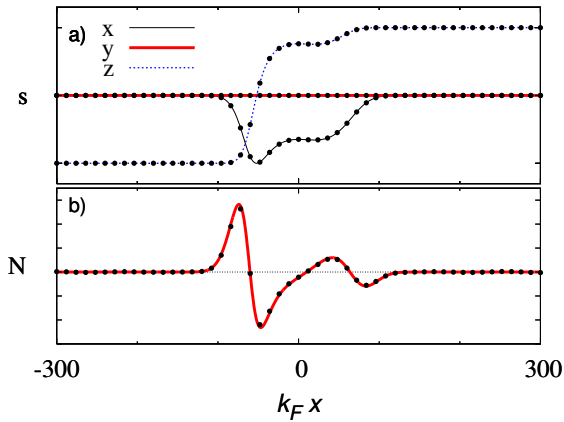}
  \caption{Equilibrium (zero-current) results: (a) Cartesian
  components of an arbitrarily chosen magnetization ${\bf M}(x)$
  (lines); Cartesian components of the calculated spin density ${\bf
  s}_{\rm eq}(x)$ (solid dots); (b) Exchange torque (solid line) and
  calculated spin-transfer torque (solid dots). }
    \label{fig:equilibrium}
\end{figure}

As a check, we used this formalism to calculate the equilibrium (zero
applied current) spin density ${\bf s}_{\rm eq}(x)$ and equilibrium
spin current density ${\bf Q}_{\rm eq}(x)$ for a magnetization
distribution chosen arbitrarily except for the constraint that $|{\bf
M}(x)|$ be uniform.  The densities ${\bf s}_{\rm eq}(x)$ and ${\bf
Q}_{\rm eq}(x)$ are obtained by retaining only the second term in
square brackets in
Eq.~(\ref{eqn:spinandcurrent}).\cite{linearBoltzmann} The lines in
Fig.~\ref{fig:equilibrium}(a) are the Cartesian components of the
imposed ${\bf M}(x)$. The solid dots in Fig.~\ref{fig:equilibrium}(a)
show that the spin density ${\bf s}_{\rm eq}(x)$ is parallel to ${\bf
M}(x)$, as expected.  Similarly, the electron-mediated spin-transfer
torque should equal the phenomenological exchange torque density
discussed by Brown.\cite{Brown:1962} This is confirmed by
Fig.~\ref{fig:equilibrium}(b), which shows that ${\bf N}_{\rm eq}(x)$
is proportional to ${\bf M}\times {\bf M}''(x)$ for the ${\bf M}(x)$
shown in Fig.~\ref{fig:equilibrium}(a).

\subsection{Semi-classical}

A semi-classical approach to calculating the spin current density is
useful for building physical intuition. Accordingly, we write an
equation of motion for the spin density of every electron that
contributes to the current. This idea has been used in the past, both
semi-quantitatively\cite{Berger:1978} and
qualitatively.\cite{Waintal:2004} Our derivation is based on the
behavior of an electron with energy $E$ that moves along the $x$-axis
through a uniform magnetic field ${\bf B}_{\rm ex}=B_{\rm ex}\hat{\bf
z}$. The wave function for such an electron is
\begin{equation}
   \varphi(x,E) = \smlb{ \begin{array}[c]{c} a~e^{ik_+ x} \\ b~e^{ik_- x} \end{array} },
\end{equation}
where 
\begin{equation}
\label{eqn:kapdef}
k_\pm^2=2mE/\hbar^2\pm k_B^2.
\end{equation}
We compute the spin density ${\bf s}(x,E)$ and the spin current
density ${\bf Q}(x,E)$ for this electron using the right sides of
Eq.~(\ref{eqn:sesd}) and Eq.~(\ref{eqn:sescd}), respectively, with
$\psi_\pm \to \varphi$.

It is straightforward to check that the components of these densities
transverse to the magnetic field satisfy the semi-classical relations
\begin{equation}
\label{eqn:scQs}
Q_x = s_x \avg{v}~~~~~~~~~~{\rm and}~~~~~~~~~~Q_y=s_y \avg{v},
\end{equation}
where $\avg{v}$ is the velocity
\begin{equation}
\label{eqn:vdef}
\avg{v}={\hbar\over m}{k_++k_-\over 2}={\hbar \over m}\avg{k}.
\end{equation}
Moreover, the transverse components of the spin density
satisfy
\begin{eqnarray}\label{eqn:perppiece}
    \avg{v}{ds_x\over dx} &=& -{\hbar k_B^2\over m}s_y,\nn
    \avg{v}{ds_y\over dx} &=& -{\hbar k_B^2\over m}s_x.
\end{eqnarray}
These equations are the components of the vector equation
\begin{equation}\label{eqn:LLGa}
    {d{\bf s}\over dx} = -{k_B^2\over \avg{k}}{\bf s}\times\hat{\bf B}_{\rm ex}
\end{equation}
where $\hat{\bf B}_{\rm ex}=\hat{\bf z}$. 

We now make the {\it ansatz} that all three Cartesian components of
the semi-classical majority and minority spin densities ${\bf
s}_\pm(x,k_x)$ satisfy Eq.~(\ref{eqn:LLGa}) when the direction of the
magnetic field varies in space. Specifically, if ${\bf B}_{\rm
ex}(x)=B_{\rm ex}\hat{\bf B}_{\rm ex}(x)$, we suppose that
\begin{equation}
\label{eqn:LLG}
    {d{\bf s}_\pm(x,k_x)\over dx} = -{k_B^2\over \avg{k}}{\bf s}_\pm(x,k_x)\times\hat{\bf B}_{\rm ex}(x),
\end{equation}
where $k_+$ and $k_-$ for ${\bf s}_+(x,k_x)$ are defined by
Eq.~(\ref{eqn:kapdef}) with $E=\hbar^2(k_x^2-k^2_{\rm
B})/2m$.\cite{KMEvan} Similarly, $k_+$ and $k_-$ for ${\bf
s}_-(x,k_x)$ are defined by Eq.~(\ref{eqn:kapdef}) with
$E=\hbar^2(k_x^2+k^2_{\rm B})/2m$. With suitable boundary conditions,
we solve the differential equation Eq.~(\ref{eqn:LLG}) to determine
the semi-classical, one-electron spin densities.
The total, spin-resolved, spin densities follow by inserting these
one-electron quantities into
\begin{equation}
\label{scm}
{\bf s}_\pm(x) 
    = \int\midb{f_\pm({\bf k}-{eE\tau\over\hbar}\bm{\hat{x}})-f_\pm({\bf k})}{\bf s}_\pm(x,k_x){k_x\over \avg{k}}d^3{\bf k}. 
\end{equation}
This equation differs from Eq.~(\ref{eqn:spinandcurrent}) by the
weighting factor $k_x/\avg{k}$.\cite{ksign} This factor guarantees
that the flux carried by each electron is proportional to its velocity
(see Appendix I). This is confirmed by Fig.~\ref{fig:LLGcheck}(b)
which shows quantitative agreement between a fully quantum calculation
of ${\bf s}(x)$ using Eq.~(\ref{eqn:Schrodinger}),
Eq.~(\ref{eqn:sesd}) and Eq.~(\ref{eqn:spinandcurrent}) and a
semi-classical calculation using Eq.~(\ref{eqn:LLG}) and
Eq.~(\ref{scm}).

\begin{figure}
  \includegraphics[width=8cm]{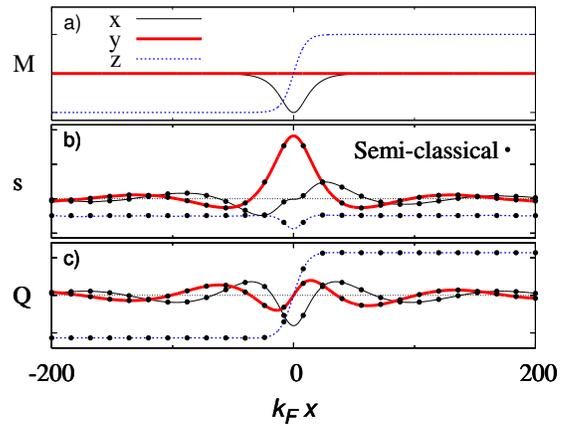}
  \caption{(a) Cartesian components of an imposed magnetization ${\bf
M}(x)$ used in the other panels; (b) comparison of quantum (lines) to
semi-classical (solid dots) calculations for the Cartesian components
of the spin density ${\bf s}(x)$; (c) same comparison for the
Cartesian components of the spin current density ${\bf Q}(x)$. }
    \label{fig:LLGcheck}
\end{figure}
In light of the foregoing, it is reasonable to calculate the
semi-classical single-electron spin current density from
\begin{equation}
\label{eqn:scscd}
{\bf Q}_\pm(x,k_x)={\bf s}_\pm(x,k_x){\hbar k_x \over m}
\end{equation}
and use the second equation in Eq.~(\ref{eqn:spinandcurrent}) to find
${\bf Q}_\pm(x)$.  The correctness of this prescription is illustrated
in Fig.~\ref{fig:LLGcheck}(c).

\section{Spin Spiral}
\label{spinspiral}

\begin{figure}
  \includegraphics[width=8cm]{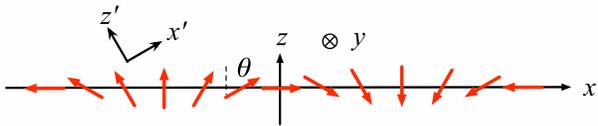}
  \caption{Cartoon of a spin spiral where the magnetization (arrows)
rotates uniformly in the $x$-$z$ plane of a fixed coordinate system
where $\hat{\bf M}(x) \cdot \hat{\bf z}=\cos\theta$. The inset shows a
local coordinate system where ${\bf M}(x)$ always points along the
$z'$ axis. }
  \label{fig:spinspiral}
\end{figure}

As a preliminary to our discussion of domain walls, it is instructive
to discuss the spin density and spin current density for a spin
spiral---an infinite magnetic structure where the direction of the
magnetization rotates continuously as one moves along a fixed axis in
space. Spin spirals occur in the ground state of some rare earth
metals\cite{Jensen:1991} and also for the $\gamma$ phase of
iron.\cite{Hafner:2002} Here, we focus on a spiral with uniform pitch
$p$ where the magnetization rotates in the $x$-$z$ plane, {i.e.}, in
Eq.~(\ref{eqn:magnetization}),
\begin{equation}
\theta(x) = p\hspace{0.1em}x ~~~~~~~~~~{\rm and}~~~~~~~~~~\phi(x)=0. 
\label{thetas}
\end{equation}
A cartoon version of this ${\bf M}(x)$ is shown in
Fig.~\ref{fig:spinspiral}. This figure also defines a {\it local}
coordinate system that will be useful in what follows. The system
$(x',y',z')$ rotates as a function of $x$ so the magnetization ${\bf
M}(x)$ always points along $+z'$.

Calvo\cite{Calvo:1978} solved Eq.~(\ref{eqn:Schrodinger}) to find the
eigenstates and eigen-energies of this spin spiral. In our notation,


\begin{equation}
    \varepsilon_{\pm}({\bf k}) 
    = {\hbar^2\over 2m}\smlb{k^2+\quarter p^2 \pm\sqrt{(k_x p)^2+k_B^4 }},
    \label{eqn:spiralenergy}
\end{equation}
and
\begin{equation}
    \psi_\pm({\bf k,r})
    = e^{i{\bf k\cdot r}} e^{-i\sigma_y\theta/2} e^{-i\sigma_x\alpha/2}\eta_\pm,
    \label{eqn:Neelstate}
\end{equation}
where
\begin{equation}
\label{alpha}
\sin\alpha = {k_x p\over \sqrt{(k_x p)^2+k_B^4}},
\end{equation}
and
\begin{equation}
    \eta_+ = \smlb{ \begin{array}[c]{c} 1 \\ 0 \end{array} },\hspace{20pt}
    \eta_- = \smlb{ \begin{array}[c]{c} 0 \\ 1 \end{array} }. \vspace{0.5em}
\end{equation}
From these results, it is easy to compute the single-electron spin
densities defined in Eq.~(\ref{eqn:sesd}). In the local $(x',y',z')$
frame,
\begin{equation}
\label{eqn:spirald}
{\bf s'}_\pm(x,k_x)=\pm(0, \sin \alpha, \cos\alpha).
\end{equation}

The corresponding calculation of the single-electron spin current
densities Eq.~(\ref{eqn:scscd}) is straightforward but tedious and not
very illuminating.  Therefore, we pass directly to the total spin
current density calculated by summing over all electrons as indicated
in the second line of Eq.~(\ref{eqn:spinandcurrent}). Again in the
local $(x',y',z')$ frame,
\begin{equation}
\label{eqn:spiralcd}
{\bf Q'}(x)=A(p,k_{\rm B})(0,0,1). 
\end{equation}
where $A(p,k_{\rm B})$ is a constant. This shows that ${\bf
Q}(x)\propto {\bf M}(x)$, {\it i.e.}, the spin current density for a
free electron spin spiral is perfectly adiabatic. Wessely {\it et
al.}\cite{Nordstrom:2005} found consistent results in their density
functional calculation of the steady-state spin current density
associated with the helical spin density wave in erbium metal.
We emphasize that  Eq.~(\ref{eqn:spiralcd}) is independent of 
pitch for an infinite spin spiral.  As we discuss below, a similar independence does {\em not}
hold for domain walls.  In that case, wide walls are adiabatic, but
narrow ones are not.

\begin{figure}
  \includegraphics[width=8cm]{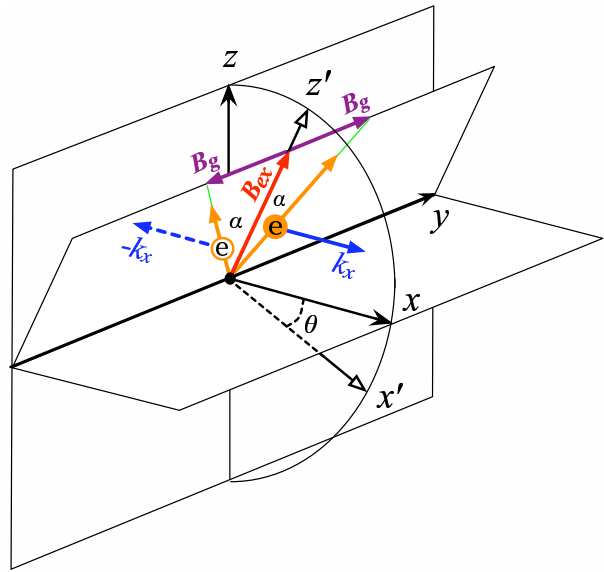}
  \caption{Electrons (wave vector $k_x$) and holes (wave vector
  $-k_x$) move in an effective field that is the sum of the exchange
  field ${\bf B}_{\rm ex}(x) \parallel \hat{\bf z}'$ and a fictitious,
  velocity-dependent ``gradient'' field induced by the spatial
  dependence of the exchange field. The spins align to the total
  effective field in an infinite spin spiral. The $x$ axis lies in
  $x'$-$z'$ plane. }
  \label{fig:curvature}
\end{figure}

The semi-classical formula Eq.~(\ref{eqn:scscd}) provides an appealing
way to understand the adiabaticity of the spin current density in the
spin spiral defined by Eq.~(\ref{thetas}). The key point is that the
angle $\alpha$ in Eq.~(\ref{alpha}) which fixes the direction of ${\bf
s}_\pm(x,k_x)$ in Eq.~(\ref{eqn:spirald}) is positive when $k_x$ is
positive and negative when $k_x$ is negative
(Fig.~\ref{fig:curvature}). Moreover, for every $k_x$ electron that
contributes to the shifted Fermi surface sums in
Eq.~(\ref{eqn:spinandcurrent}), there is a contribution from a $-k_x$
hole. Now, a hole has opposite spin density to an electron and the
spin current density Eq.~(\ref{eqn:scscd}) contains an additional
factor of $k_x$. Therefore, the two spin density vectors in
Fig.~\ref{fig:curvature} {\it subtract} to give a net spin density
along $\hat{\bf y}$ while the corresponding two spin current density
vectors {\it add} to give a net spin current density along $\hat{\bf
z}'$.  This occurs for all $k_x$ in the sums so ${\bf Q}(x)$ aligns
exactly with the local exchange field and thus with the local
magnetization.

The opposite situation occurs for fully occupied states below the
Fermi energy.  The spins of the forward and backward moving electrons
combine to produce a net moment aligned with the exchange field, as
necessary for self-consistency.  Further, the spin currents, with the
additional factor of $k_x$ add to give a net spin current along
$\hat{\bf y}$, so that its gradient gives the correct form of the
phenomenological exchange torque density.

To summarize: an electric current that passes through a spin spiral
generates a spin accumulation with a component transverse to the
magnetization. The spin current density possesses {\it no} such
component due to pairwise cancellation between forward and backward
moving spins of the same type (majority or minority). Moreover, since
the cancellation occurs within each band, the final result is
insensitive to the details of intraband scattering.

It remains only to understand the origin of the misalignment angle
$\alpha$. Why does each spin not simply align itself with ${\bf
B}_{\rm ex}$?  Berger\cite{Berger:1978} was the first to notice this
fact and the physics was made particularly clear by Aharonov and
Stern.\cite{Aharonov:1992} These authors studied the adiabatic
approximation for a classical magnetic moment that moves in a slowly
varying field ${\bf B}(x)$. Not obviously, the moment behaves as if
were subjected to a effective magnetic field ${\bf B}_{\rm eff}(x)$
that is the sum of ${\bf B}_{\rm ex}(x)$ and a fictitious, velocity-dependent, ``gradient''
field ${\bf B}_{\rm g}(x)$ that points in the direction $\nabla
\hat{\bf B}(x)\times \hat{\bf B}(x)$. For our problem,
\begin{equation}
\label{eqn:gradientfield}
    {\bf B}_{\rm eff} 
    = {\bf B}_{\rm ex}+{\hbar^2k_x\over 2m\mu}{d\hat{\bf B}_{\rm ex}\over dx}\times\hat{\bf B}_{\rm ex}. 
  \end{equation}
The presence of this gradient field is apparent from
Eq.~(\ref{eqn:spiralenergy}) where the square root is proportional to
$\vert {\bf B}_{\rm eff}\vert$. {\it The adiabatic solution
corresponds to perfect alignment of the moment with ${\bf B}_{\rm
eff}(x)$}.  This alignment is indicated in Eq.~(\ref{eqn:spirald}) and
in Fig.~\ref{fig:curvature}.  More generally, the expected motion of
the magnetic moment is precession around ${\bf B}_{\rm
eff}(x)$.Nevertheless, as indicated above, the total
spin current density for the spin spiral aligns with ${\bf B}_{\rm
ex}(x)$ (which is the conventional definition of adiabaticity for this
quantity) when the net effect of all conduction electrons is taken
into account.





\section{Domain Walls}
\label{domainwalls}
Our main interest is the spin-transfer torque associated with domain
walls that connect two regions of uniform and antiparallel
magnetization. A realistic wall of this kind can be described by
Eq.~(\ref{eqn:magnetization}) with\cite{Hubert:1998}
\begin{equation}
\label{eqn:asine}
\theta(x) = \pi/2 - \arcsin\midb{\tanh\smlb{x/w}} ,
\end{equation}
The wall is N\'eel-type if $\phi(x)=0$ and Bloch-type if
$\phi(x)=\pi/2$.  We will speak of the domain wall width $w$ as
``long'' or ``short'' depending on whether $w$ is large or small
compared to the characteristic length
\begin{equation}
\label{eqn:length}
L={E_F\over E_{\rm ex}}{1\over k_F}={k_F\over k_B^2}.
\end{equation}

\begin{figure}
 \includegraphics[width=8cm]{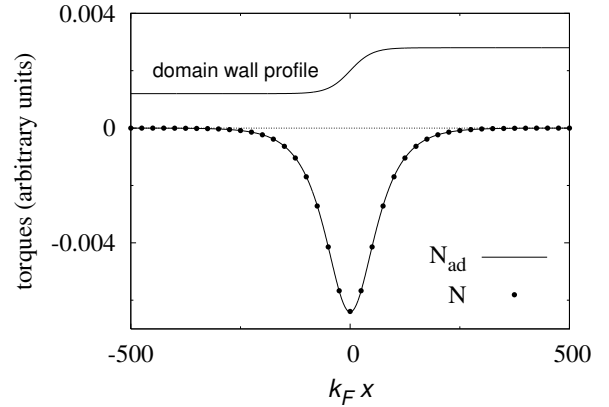}
 \caption{Distributed spin-transfer torque for a long N\`eel domain
 wall with $w=50$ and $L=6.25$ ($k_F=1$ and $k_B=0.4$): semi-classical
 calculation of ${\bf N}_{\rm st}$ (solid dots) compared to
 Eq.~(\ref{eqn:adit}) for ${\bf N}_{\rm ad}(x)$ (solid curve).}
 \label{fig:longwalltorque}
\end{figure}

Intuitively, the adiabatic approximation should be valid when $w \gg
L$.  When applied to Eq.~(\ref{eqn:spinandcurrent}), the predicted
adiabatic spin-transfer torque for our model is
\begin{equation}
\label{eqn:adit}
{\bf N}_{\rm ad}(x) = -{\hbar\over 2}\eta{neE\tau\over m}\partial_x\hat{\bf M}(x),
\end{equation}
where $n$ is the electron density, and $\eta$ is the polarization of
the current.  The calculations required to check this for long domain
walls are difficult quantum mechanically (for numerical reasons) but
straightforward semi-classically.  At the single-electron level,
adiabaticity again corresponds to alignment of the spin moment with
the effective field defined in Eq.~(\ref{eqn:gradientfield}).  The
results for a typical long wall (Fig.~\ref{fig:longwalltorque})
demonstrate that summation over all electrons produces alignment of
${\bf Q}(x)$ with ${\bf M}(x)$ so the adiabatic formula
Eq.~(\ref{eqn:adit}) is indeed correct in this limit.

For short walls, we have carried out calculations of ${\bf N}_{\rm st}(x)$ both
quantum mechanically and semi-classically. The two methods agree very
well with one another (see Fig.~\ref{fig:LLGcheck}) but not with the
proposed form Eq.~(\ref{lt}). Bearing in mind that, when the
magnetization changes, $\hat{\bf x}'$ points along $\partial_x {\bf
M}$ and $\hat{\bf y}$ points along ${\bf M}\times\partial_x{\bf M}$,
our result for the spin-transfer torque is
\begin{equation}\label{eqn:Tdecompose}
   {\bf N}_{\rm st}(x) = {\bf N}_{\rm ad}(x)+a(x)\hat{\bf x}' + b(x)\hat{\bf y}.
\end{equation}

\begin{figure}
 \includegraphics[width=8cm]{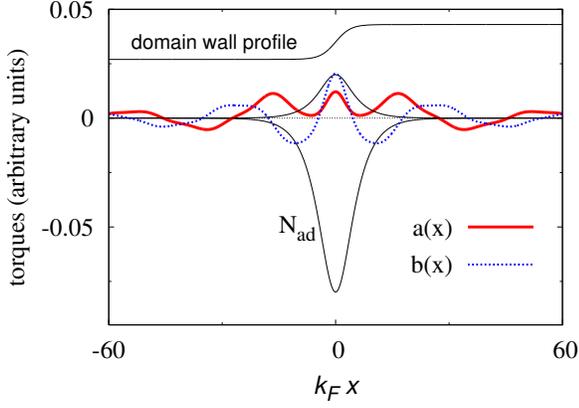}
 \caption{Distributed spin-transfer torque for a short N\`eel domain
 wall with $w=4$ and $L=6.25$ ($k_F=1$ and $k_B=0.4$): in-plane piece
 $a(x)$ (heavy solid curve); out-of-plane piece $b(x)$ (dashed curve);
 adiabatic prediction (light solid curve); second term in
 Eq.~(\ref{lt}) scaled to match the maximum of $b(x)$ (light solid
 curve). }
 \label{fig:shortwalltorque}
\end{figure}

${\bf N}_{\rm st}(x)$ differs from ${\bf N}_{\rm ad}(x)$ because gradients in
the gradient field induce single electron spin moments to precess
around ${\bf B}_{\rm eff}(x)$ rather to align perfectly with
it. Fig.~\ref{fig:shortwalltorque} shows $a(x)$ and $b(x)$ as
calculated for a typical short domain wall. The associated torques lie
in the plane of the magnetization and perpendicular to that plane,
respectively. These non-adiabatic contributions to the torque are both
oscillatory functions of position that do not go immediately to zero
when the the magnetization becomes uniform. In other words, $a(x)$ and
$b(x)$ are generically non-local functions of the magnetization ${\bf
M}(x)$.
The positive-valued function that falls to zero at the edges of the
domain wall (light solid curve in Fig.~\ref{fig:shortwalltorque}) is
the second function in Eq.~(\ref{lt}) with $c_2$ chosen to match
$b(x)$ at their common maximum.  Evidently, the proposed torque
function Eq.~(\ref{lt}) gives at best a qualitative account of the
out-of-plane non-adiabatic torque.

A convenient measure of the degree of non-adiabaticity of the
spin-transfer torque is
\begin{equation}
\label{eqn:nan}
{\cal Q}= {{\rm max}\,|\,b(x)\,| \over {\rm max}\,|\,N_{\rm ad}(x)\,|}.
\end{equation}
\begin{figure}
 \includegraphics[width=8cm]{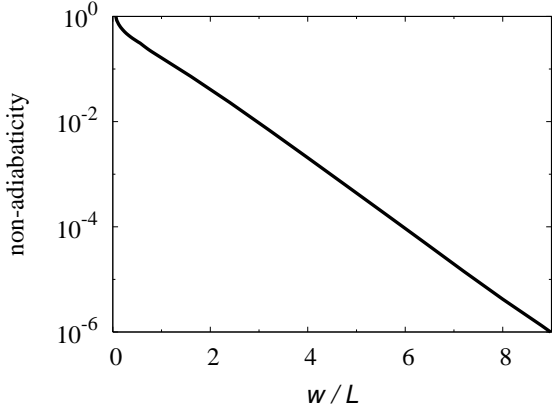}
 \caption{Non-adiabaticity in Eq.~\ref{eqn:nan} versus wall width
 scaled by the characteristic length in Eq.~(\ref{eqn:length}). Note
 the logarithmic scale.}
   \label{fig:nonadiabaticity}
\end{figure}
Fig.~\ref{fig:nonadiabaticity} plots this quantity as a function of
scaled domain wall width $w/L$ on a log scale. The observed
exponential decrease of the non-adiabatic torque as the wall width
increases can be understood from the work of Dugaev {\it et
al.}\cite{Dugaev:2002} These authors treat the gradient field in
Eq.~(\ref{eqn:gradientfield}) as a perturbation and calculate the
probability for an electron in a $(k_x \uparrow)$ state to scatter
into a $(k'_x \downarrow)$ state in the Born approximation. If we
choose $k_x$ and $k'_x$ as $k_F^+$ and $k_F^-$, respectively, their
results imply that the probability ${\cal P}$ that a majority electron
retains its spin and becomes a minority electron as it passes through
a domain wall is
\begin{equation}
\label{eqn:Dugaev}
{\cal P} \propto \exp(-\gamma w/L),
\end{equation}
where $\gamma$ is a constant of order unity. This rationalizes the
result plotted in Fig.~\ref{fig:nonadiabaticity} because the magnitude
of the minority spin component determines the amplitude of the spin
precession around ${\bf B}_{\rm eff}(x)$ and thus the magnitude of the
non-adiabatic component of ${\bf s}$ and ${\bf Q}$ in
Eq.~(\ref{eqn:scscd}). In fact, $N_{\rm ad} \propto 1/w$, so it is the
case that

\begin{equation}
\label{over}
 {\rm max}\,|\,b(x)\,|\propto {1\over w} \exp(-\gamma w/L).
\end{equation} 

\begin{figure}
 \includegraphics[width=8cm]{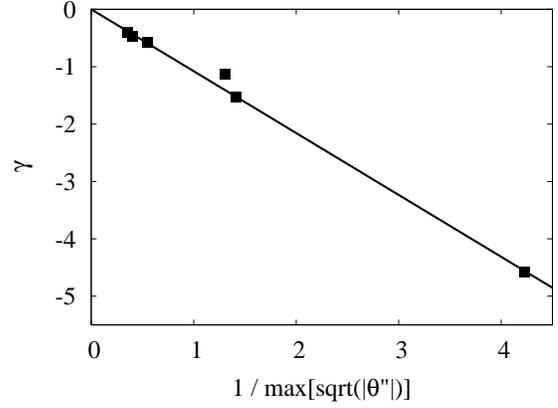}
 \caption{Dependence of $\gamma$ in Eq.~(\ref{eqn:Dugaev}) on domain
 wall sharpness: Squares are calculated points. Straight line is a
 guide to the eye that passes through the origin.}
   \label{fig:slope}
\end{figure}

The slope of the straight line in Fig.~\ref{fig:nonadiabaticity}, {\it
i.e.}, the value of the constant $\gamma$ in Eq.~(\ref{eqn:Dugaev})
depends on the sharpness of the domain wall. Using
Eq.~(\ref{eqn:asine}) and other simple domain wall profile functions,
it is not difficult to convince oneself that a suitable measure of
domain wall sharpness is the maximum value of the second derivative
$\theta''(x)$ for walls with the same width. The numerical results
shown in Fig.~\ref{fig:slope} confirm this to be true. The sharper the
domain wall, the less rapidly the non-adiabatic torque disappears with
increasing domain wall width.
 
\section{Relation to Other Work}
\label{other}
\subsubsection{Waintal \& Viret}

Waintal and Viret\cite{Waintal:2004} (WV) used a free-electron Stoner
model and the Landauer-B\"{u}ttiker formalism to calculate the spin
transfer torque associated with a N\'eel wall with magnetization
Eq.~(\ref{eqn:magnetization}) and
\begin{eqnarray}
\label{WVwall}
  \quad
   \theta(x) &=& \left\{ \begin{array}{lr}
   0, & x < -w \\
   \displaystyle (\pi/2)\smlb{x/w + 1}, & -w\le x\le w \\
   \pi, & x > w .
   \end{array}\right.  \nn
\end{eqnarray}
For this wall profile (which is exactly one half-turn of a uniform
spin spiral in the interval $-w\le x\le w$), WV reported oscillatory
non-adiabatic contributions to the torque similar to our functions
$a(x)$ and $b(x)$. This contrasts with the perfect adiabaticity we
found in Sec.~\ref{spinspiral} for the infinite spin spiral. Moreover,
the amplitude of the non-adiabatic torque reported by WV for this wall
decreases only as $1/w$ rather than $(1/w)\exp(-\gamma w/L)$ as we
found above.

The disparities between Ref.~\onlinecite{Waintal:2004} and the present
work all arise from the unphysical nature of the domain wall
Eq.~(\ref{WVwall}). Specifically, the divergence of $\theta''(x)$ at
$x=\pm w$ locates this wall at the origin of Fig.~\ref{fig:slope}
where $\gamma=0$. This brings their result into agreement with
Eq.~(\ref{over}).  Any rounding of the discontinuity in slope at
$x=\pm w$ would yield a finite value for $\theta''(x)$ and thus a
non-zero value of $\gamma$.

In Appendix II, we calculate the spin-transfer torque for the wall
Eq.~(\ref{WVwall}) using our methods.  Qualitatively, the  pure $1/w$
behavior of the non-adiabatic torque comes  from the fact that there
is a sudden jump in $\theta'(x)$ at $x=\pm w$. There is a
corresponding jump in the direction of ${\bf B}_{\rm eff}(x)$ as
defined by Eq.~(\ref{eqn:gradientfield}). Spins propagating along the
the $x$-axis cannot follow this abrupt jump and thus precess around
the post-jump field direction with an amplitude determined by the sine
of the angle between the before-and-after field directions. The latter
is proportional to the jump in $\theta'(x)$, which is $\pi/2w$ for the
wall Eq.~(\ref{WVwall}). 

\subsubsection{Zhang \& Li}

In spin spirals and long domain walls, we find that the
non-equilibrium spin current is adiabatic, {\it i.e.}, ${\bf Q}(x)$ is
aligned with ${\bf M}(x)$ [or ${\bf B}_{\rm ex}(x)$].  At the same
time, we find in both cases that the non-equilibrium spin density
${\bf s}(x)$ is {\em not} aligned with the magnetization; there is
component of ${\bf s}(x)$ transverse to ${\bf M}(x)$. The
corresponding transverse component of the spin current density cancels
between pairs of electrons moving in opposite directions.  Zhang and
Li\cite{Zhang:2004} found exactly the same form of non-equilibrium
spin accumulation (called $\delta {\bf m}(x)$ by them) using a
phenomenological theory. They proposed that this non-equilibrium spin
density relaxes by spin-flip scattering toward alignment with the
magnetization. Such a relaxation would produce a non-adiabatic torque
of the form given by the second term in Eq.~(\ref{lt}).  The
correctness of this predicted non-adiabatic torque depends on the
correctness of the assumed model for relaxation of transverse spin
accumulation through spin flip scattering.

Zhang and Li assume a form for the rate of spin flip scattering, $\delta {\bf m}
/\tau_{\rm sf}$, that has been used successfully as a phenomenological
description of {\it longitudinal} spin relaxation in systems with collinear
magnetization.  While it is plausible to extend this form,
as they do, to describe {\it transverse} spin relaxation in non-collinear
systems, our calculations indicate that it is not likely to be
correct.  Our reasoning is simplest to appreciate for a spin spiral
with small pitch $p$. In this limit, Eqs.~(\ref{alpha})~and~(\ref{eqn:spirald}) show that the transverse component of the spin for every electron eigenstate is proportional to its velocity. This means that the majority band electrons contribute a transverse spin accumulation and an electric current that are proportional to one other. The same is true, separately, for the minority band electrons.  This conclusion is independent of the details of the electron distribution.  Therefore, for a fixed total current, it is impossible to relax the transverse spin accumulation without changing the
longitudinal polarization of the current.  No such change occurs in
the model in Ref.~\onlinecite{Zhang:2004}, casting doubt on the validity
of the form of the spin flip scattering assumed there.

Microscopic considerations also argue against this form of the
relaxation.  As we have emphasized, the natural basis for an electron
spin moving though a non-collinear magnetization is not along the
local exchange field ${\bf B}_{\rm ex}(x)$, but rather along a local
effective field ${\bf B}_{\rm eff}(x)$, which includes the corrections
due to the gradient of the magnetization [see
Eq.~(\ref{eqn:gradientfield})].  Any spin that deviates from parallel
or antiparallel alignment with the effective field will precess around
the effective field, and on average will point parallel or
antiparallel.  Thus, we expect that there is {\it no} tendency for
electron spins moving in a non-uniform magnetization to align
themselves with the local exchange field ${\bf B}_{\rm ex}(x)$ by
spin-flip scattering (or any other mechanism).  Rather, the adiabatic
solution is precisely alignment of their spins with the local
effective field ${\bf B}_{\rm eff}(x)$.  Without further microscopic
justification, we believe that the phenomenological form of spin flip
scattering assumed in Ref.~\onlinecite{Zhang:2004} should not be used
in systems with non-collinear magnetizations.  Hence, this analysis
argues against the existence of the resulting contribution to the
``non-adiabatic'' torque from spin flip scattering.


\section{Scattering}
\label{scattering}
We do not explicitly treat scattering in any of our calculations.
However, the distribution function in Eq.~(\ref{eqn:spinandcurrent}),
a shifted Fermi distribution, is an approximate solution of the
Boltzmann equation in certain limits.  First, the electric field must
be small enough that the transport is in the linear regime.  Then, the
appropriate limits are determined by three important length scales,
the Fermi wavelength, the mean free path, and the characteristic
length of the structure, either the pitch of the spin spiral or the
width of the domain wall.  In all cases, we consider the limit in
which the Fermi wavelength is short compared to the mean free path.
This limit allows the description of the states of the system in terms
of the eigenstates of the system in the absence of scattering.
Different limits apply to the cases of domain walls and of spin
spirals because the distribution functions are interpreted differently
for these two structures.

We use the Boltzmann equation in two different ways.  When the mean free
path is much longer than the characteristic size of the structure, the
distribution function describes the occupancy of the eigenstates of the
entire system.  This distribution function is independent of the spatial
coordinate and we refer to this approach as global.  In the opposite
limit, the distribution function is spatially varying and describes the
occupancy of eigenstates of the local Hamiltonian, which includes the
exchange field and the gradient field.  We refer to this approach as
local, as the distribution function can vary spatially.

For spin spirals, the distribution functions are shifted Fermi
functions of the eigen-energies of the spin spiral.  In the limit that
the pitch of the spiral is much shorter than the mean free path, the
shifted distribution given in Eq.~(\ref{eqn:spinandcurrent}) is a
solution of the global Boltzmann equation in the relaxation time
approximation.  The distribution function also becomes a solution in
the opposite limit, where the mean free path is much shorter than the
pitch of the spiral.  In this limit, the Boltzmann equation is
considered locally rather than globally.  At each point in space the
states are subject to the local exchange field, and the local gradient
field.  The distribution function is defined for states that are
locally eigenstates of the sum of the fields.  The local distribution
function is given by the adiabatic evolution in the rotating reference
frames of the distribution function specified in
Eq.~(\ref{eqn:spinandcurrent}). In the limit that the pitch of the
spiral goes to infinity, this distribution function locally solves the
Boltzmann equation in the relaxation time approximation.  Thus, for
spin spirals, the distribution function given in
Eq.~(\ref{eqn:spinandcurrent}) is a solution in the limits that the
mean free path is much greater than or much less than the pitch.  We
speculate that the corrections in between these limits are small.

Domain walls are not uniform in the way that spin spirals are, so the
distribution functions need to be given a different interpretation.
For these structures, the distribution function is determined from the
properties of the states in the leads. For example, in the
Landauer-B\"{u}ttiker approach to this problem,\cite{Waintal:2004}
scattering is ignored in the domain wall itself and confined to the
``leads'' adjacent to it (these leads are assumed to be ``wide'' and
function as electron reservoirs). An applied voltage is assumed to
raise the energy of electron states in one lead relative to the
other. Thus, in a formula like Eq.~(\ref{eqn:spinandcurrent}), the
distribution function is shifted in energy rather than in velocity.

We also do not treat scattering within the domain wall explicitly, but
we assume that the wall is bounded by long leads that are as
``narrow'' as the domain wall region and have resistances per unit
length that are comparable to that of the domain wall region.  Thus,
the distribution of the states approaching the domain wall region is
similar to the distribution of states in an extended wire, {\it i.e.},
to that given by Eq.~(\ref{eqn:spinandcurrent}).  For domain walls in
long wires, the distribution function for left going states is
determined by the right lead and for right going states by the left
lead.  With this interpretation, the distribution given in
Eq.~(\ref{eqn:spinandcurrent}) is a solution in the limit that the
scattering in the domain wall is weak, that is, the domain wall is
much narrower than the mean free path.

The distribution in Eq.~(\ref{eqn:spinandcurrent}) is also a solution
in the limit that the mean free path is much shorter than the domain
wall width.  Since the Fermi wave length is much shorter than the mean
free path, it is much less than the domain wall width.  In this case,
quantum mechanical reflection is negligible and the quantum mechanical
states are closely related to the semiclassical trajectories.  With a
similar interpretation of the distribution function as was made for
the spin spirals in this limit, the same conclusion holds for the
domain walls.

\section{Summary \& Conclusion}
\label{sumcon}

In this paper, we analyzed spin-transfer torque in systems with
continuously variable magnetization using previous results of
Calvo\cite{Calvo:1978} for the eigenstates of an infinite spin spiral
and of Aharonov and Stern\cite{Aharonov:1992} for the classical motion
of a magnetic moment in an inhomogeneous magnetic field. Adiabatic
motion of individual spins corresponds to alignment of the spin moment
{\it not} with the exchange field (magnetization) but with an
effective field that is slightly tilted away form the exchange field
by an amount that depends on the spatial gradient of the
magnetization. Nevertheless, when summed over all conduction
electrons, the spin current density is parallel to the magnetization
both for an infinite spin spiral and for domain walls that are long
compared to a characteristic length $L$ that depends on the exchange
energy and the Fermi energy.

Non-adiabatic corrections to the spin-transfer torque occur only for
domain walls with widths $w$ that are comparable to or smaller than
$L$. The non-adiabatic torque is oscillatory and non-local in space
with an amplitude that decreases as $w^{-1}\exp(-\gamma w/L)$. The
constant $\gamma$ is largest for walls with the sharpest magnetization
gradients. This suggests that non-adiabatic torques may be important
for spin textures like vortices where the magnetization varies
extremely rapidly.

Using microscopic considerations, we
have also argued that the role of the gradient field to tilt spins away from the exchange
field casts serious doubt on a recent proposal by Zhang and
Li\cite{Zhang:2004} that a non-negligible non-adiabatic contribution
to the torque arises from relaxation of the non-equilibrium spin
accumulation to the magnetization vector by spin flip scattering.
We conclude that, if the second term in Eq.~(\ref{lt}) truly
accounts for the systematics of current-driven domain wall motion, the
physics that generates this term still remains to be identified.

Finally, we have carefully discussed the role of scattering in this
problem with particular emphasis on the approximation used here to
neglect scattering within the domain wall itself but to treat the
adjacent ferromagnetic matter as bulk-like. We argue that this
approximation is valid in limits that either include or bracket the
most interesting experimental situations and therefore is likely to be
generally useful.

\section{Acknowledgment}
One of us (J.X.) is grateful for support from the Department of Energy
under grant DE-FG02-04ER46170. 
\vspace{1em}
\begin{center}
  {\bf Appendix I: Semi-Classical Weighting Factor}
\end{center}
The weighting factor $k_x/\avg{k}$ used in Eq.~(\ref{scm}) brings the
amplitude of the dynamic (transverse) part of the semi-classical,
one-electron spin density into accord with the corresponding quantum
mechanical amplitude. This can be seen from a simple model problem
that we solve both quantum mechanically and semi-classically. Namely,
a spin initially oriented along the $+\hat{\bf x}$ direction
propagates from $x=-\infty$ to $x=\infty$ through a magnetization that
changes abruptly from ${\mathbf M}(x) = M(1,0,0)$ for $x<0$ to
${\mathbf M}(x) = M(0,0,1)$ for $x \ge 0$.  For $x<0$, the eigenstates
are
\begin{equation}
   \psi_\up^-(x) = {1\over\sqrt{2}} \smlb{ \begin{array}[c]{c} 1 \\ 1 \end{array} } e^{ik_\up x},\quad
   \psi_\down^-(x) = {1\over\sqrt{2}}\smlb{ \begin{array}[c]{c} 1 \\ -1 \end{array} } e^{ik_\down x},
\end{equation}
and for $x>0$, the eigenstates are
\begin{equation}
  \psi_\up^+(x) = \smlb{ \begin{array}[c]{c} 1 \\ 0 \end{array} } e^{ik_\up x},\quad
  \psi_\down^+(x) = \smlb{ \begin{array}[c]{c} 0 \\ 1 \end{array} } e^{ik_\down x}.
\end{equation}

If we choose the incoming state as
\begin{equation}
  \psi(x) = \psi_\up^-(x),
\end{equation}
the reflection and transmission amplitudes for spin flip
$(r_{\up\down}, t_{\up\down})$ and no spin flip
$(r_{\up\up},t_{\up\up})$ are determined by matching the total wave
function and its derivative at $x=0$:
\begin{eqnarray}
  \psi_\up^- + r_{\up\up} (\psi_\up^-)^* + r_{\up\down} (\psi_\down^-)^* &=& t_{\up\up} \psi_\up^+ + t_{\up\down} \psi_\down^+  \nn
&& \\
  k_\up\psi_\up^- - r_{\up\up} k_\up(\psi_\up^-)^* - r_{\up\down} k_\down(\psi_\down^-)^* &=& t_{\up\up} k_\up\psi_\up^+ + t_{\up\down} k_\down\psi_\down^+. \nn
\end{eqnarray}
It is straightforward to confirm that these equations are solved by
\begin{eqnarray}
   r_{\up\up} = {k_\up^2-k_\down^2\over k_\up^2 + 6k_\up k_\down  + k_\down^2} &,&
   r_{\up\down} = {2k_\up(k_\down-k_\up)\over k_\up^2 + 6k_\up k_\down + k_\down^2} \nn
   t_{\up\up} = {4\sqrt{2} k_\up k_\down\over k_\up^2 + 6k_\up k_\down  
+ k_\down^2} &,&
   t_{\up\down} = {2\sqrt{2}k_\up(k_\up+k_\down)\over k_\up^2 + 6k_\up k_\down + k_\down^2}. \nn
\end{eqnarray}
We are interested in the transmitted wave function,
\begin{equation}
  \psi_{\rm tr}(x) = t_{\up\up}\psi_\up^+(x) + t_{\up\down}\psi_\down^+(x)
  = \smlb{ \begin{array}[c]{c} t_{\up\up}e^{ik_\up x} \\ t_{\up\down} e^{ik_\down x} \end{array} },
\end{equation}
which carries a spin density,
\begin{equation}
\label{ampq}
  {\mathbf s}_{\rm tr}^{\rm qm}(x) = {\hbar\over 2} \left[2t_{\up\up}t_{\up\down}\cos(\delta kx), 2t_{\up\up}t_{\up\down}\sin(\delta kx), (t_{\up\up}^2-t_{\up\down}^2)\right],
\end{equation}
where $\delta k=k_\up - k_\down$. Notice that the oscillation
is transverse to the $x\to\infty$ magnetization and of amplitude
$\hbar t_{\up\up}t_{\up\down}$.

If we analyze the same problem semi-classically, a majority electron
propagates freely until it reaches $x=0$. At that point, the electron
feels a magnetization perpendicular to its magnetic moment and begins
precession around that magnetization with unit amplitude. The
associated spin density is
\begin{equation}
\label{ampsc}
   {\mathbf s}_{\rm tr}^{\rm sc}(x) = {\hbar\over 2} \left [\cos(\delta kx),\sin(\delta kx),0\right ].
\end{equation}

\begin{figure}
\includegraphics[width=8cm]{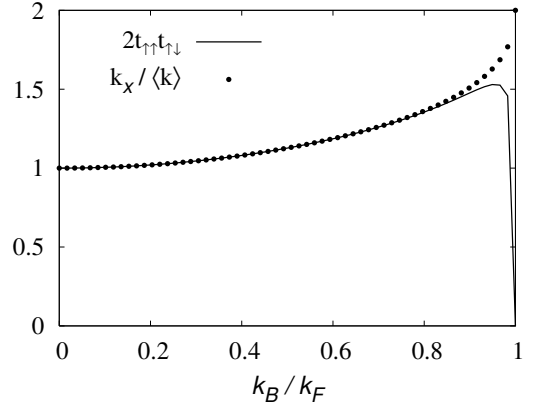}
\caption{The semi-classical weighting factor for the spin
density. Solid (dotted) curve is the expression on the left (right)
side of the $\approx$ symbol in Eq.~(\ref{eqn:approxa}). }
\label{fig:scfactor}
\end{figure}

Comparing Eq.~(\ref{ampq}) to Eq.~(\ref{ampsc}) shows that the
transverse oscillation amplitudes will be equal if we multiply the
semi-classical result by the weighting factor
\begin{equation}
\label{eqn:approxa}
  2t_{\up\up}t_{\up\down} =
  {32 k_\up^2 k_\down(k_\up+k_\down)\over (k_\up^2 + 6k_\up k_\down  + k_\down^2)^2}
  \approx {2k_\up\over k_\up+k_\down} = {k_x\over \avg{k}}.
\end{equation}
Fig.~\ref{fig:scfactor} illustrates the quality of the approximation
in Eq.~(\ref{eqn:approxa}) if we identify $k_\up$ and $k_\down$ with
$k^+_F$ and $k^-_F$ (respectively) in Eq.~(\ref{eqn:kpm}). Of course,
$k_x$ plays the role of $k_\up$ in Eq.~(\ref{scm}).

\vspace{1em}
\begin{center}
  {\bf Appendix II: Spin Spiral Domain Wall}
\end{center}
The semi-classical spin density associated with electron propagation
through a magnetization like Eq.~(\ref{eqn:magnetization}) is the
solution of Eq.~(\ref{eqn:LLG}) with suitable boundary
conditions. Choosing $\phi=0$, we simplify the notation by using the
prefactor $\lambda=k_B^2/\avg{k}$ and an overdot for $d/dx$ to write
the components of Eq.~(\ref{eqn:LLG}) as
\begin{eqnarray}
  \dot{s}_x &=& - \lambda s_y\cos\theta \\
  \dot{s}_y  &=& - \lambda s_z\sin\theta + \lambda s_x\cos\theta \\
  \dot{s}_z  &=& \lambda s_y\sin\theta.
\end{eqnarray}
In the local frame $(x',y',z')$ defined in Fig.~\ref{fig:spinspiral},
the components of the spin density,
\begin{eqnarray}
  s'_x &=& s_x\cos\theta - s_z\sin\theta \\
  s'_y &=& s_y \\
  s'_z &=& s_x\sin\theta + s_z\cos\theta,
\end{eqnarray}
satisfy 
\begin{eqnarray}
\dot{s}'_x  &=& -\lambda s'_y - s'_z\dot{\theta} \\
\dot{s}'_y &=& \lambda s'_x \\
\dot{s}'_z  &=& s'_x\dot{\theta}.
\end{eqnarray}
Eliminating $s'_y$ gives
\begin{equation}
  \ddot{s}'_x + (\lambda^2+\dot{\theta}^2)s'_x + s'_z\ddot{\theta} = 0.
  \label{eqn:mX}
\end{equation}

The differential Eq.~(\ref{eqn:mX}) cannot be solved analytically for
realistic domain wall profiles. However, it is easily solvable for the
wall defined by Eq.~(\ref{WVwall}) where one-half turn of a spin
spiral with pitch $p=\pi/2w$ connects two regions with uniform (but
reversed) magnetization. In the limit $\pi/w\ll\lambda$ of a long
wall, the components of the spin density transverse to the wall
magnetization for the range $x\in[-w,w]$ are (after multiplying the
weighting factor $k_x/\avg{k}$ for the semi-classical approach)
\begin{eqnarray}
   s'_x(x) &=& {\hbar\over 2}{k_x\over \avg{k}}{\pi \over 2w\lambda}\sin(\lambda(x\pm w)) \nn
   s'_y(x) &=& {\hbar\over 2}{k_x\over \avg{k}}{\pi \over 2w\lambda}\midb{1-\cos(\lambda(x\pm w))},
\end{eqnarray}
where the plus (minus) refers to electrons that flow from left (right)
to right (left). The associated spin current density and spin transfer
torque carried by each electron follow from Eq.~(\ref{eqn:scscd}) and
Eq.~(\ref{eqn:spt}), respectively.  Bearing in mind that $\hat{\bf
x}'$ varies with $x$, our final result for the torque (in the local
frame) generated by a single electron moving from right to left is
\begin{eqnarray}
  N'_x &=& {\hbar\over 2}{\hbar k_x\over m}{k_x\over \avg{k}}{\pi\over 2w}\left[1-\cos\lambda(x-a)\right]~\hat{\bf x}' \nn
  N'_y &=& {\hbar\over 2}{\hbar k_x\over m}{k_x\over \avg{k}}{\pi\over 2w}\sin\lambda(x-a)~\hat{\bf y}.
\end{eqnarray}
This may be compared with the results of
Ref.~\onlinecite{Waintal:2004} which pertain to the entire ensemble of
conduction electrons.




\begin{thebibliography}{99}
\bibitem{MagSign}
In reality, the magnetization is oppositely directed to the spin
density since the g-factor for electrons is negative.
For simplicity, we ignore this sign difference and assume that the
magnetization is parallel to the spin density.  Care is required when
using spin currents to compute torques on magnetizations.
\bibitem{linearBoltzmann} 
When $E\tau$ is sufficiently small, as we
assume here, the difference in the square brackets in
Eq.~(\ref{eqn:spinandcurrent}) is zero except in the immediate
vicinity of the Fermi surface.  In this limit, the integral over
reciprocal space in Eq.~(\ref{eqn:spinandcurrent}) reduces to an
integral over wave vectors restricted to the Fermi surface, reflecting
the fact that non-equilibrium transport involves states near the Fermi
surface.  The reduction of the integration in this manner leads to the
linearized Boltzmann equation.  On the other hand, the equilibrium
spin densities and currents, as shown in Fig.~\ref{fig:equilibrium},
involve contributions from all occupied states, not just those near
the Fermi surface.
\bibitem{KMEvan} 
    For $k_x^2<2k_B^2$, otherwise $k_-=0\Ra \avg{k}=k_+/2$, which ensures that $\avg{k}$ is continous function of $k_x$.
\bibitem{ksign} We note that $k_x$ and $\avg{k}$ have the same algebraic sign.
\end{thebibliography}
\end{document}